\documentstyle[aps,epsf]{revtex}
\begin{document}
\draft
\twocolumn[\hsize\textwidth\columnwidth\hsize\csname 
@twocolumnfalse\endcsname
\title{Generation and Structure of Solitary Rossby Vortices in 
Rotating Fluids}
\author{Nikolai Kukharkin\cite{em} and Steven A.\ Orszag}
\address{Fluid Dynamics Research Center,
Princeton University, Princeton, New Jersey 08544}
\date{\today}
\maketitle
\begin{abstract}
The formation of zonal flows and vortices in the generalized
Charney-Hasegawa-Mima equation is studied. We focus on the regime when the
size of structures is comparable to or larger than the deformation (Rossby)
radius.  Numerical simulations show the formation of anticyclonic vortices in
unstable shear flows and ring-like vortices with quiescent cores and
vorticity concentrated in a ring. Physical mechanisms that lead to these
phenomena and their relevance to turbulence in planetary atmospheres are
discussed. \end{abstract}
]

Fluid motion subject to strong rotation, stratification, or action of a
magnetic field can often be considered as quasi-two-dimensional. One of the
major features of 2D flows is their tendency to self-organization which
reveals itself by spontaneous generation of coherent structures (e.g.
vortices, jets) that dominate the large-scale motion. The dynamics of these
structures becomes more complicated when the inverse energy cascade
interferes with the characteristic spatial scale determined by the external
field. The presence of the Coriolis force due to rotation ($\beta$-effect)
leads to a limitation on vortex sizes in the meridional direction by the
Rhines length \cite{rh}, while in the strong turbulence regime when waves are
neglected, the influence of the finite deformation (Rossby) radius results in
the emergence of ``shielded'' vortices, which can form a quasicrystalline
structure \cite{prl}. 

In this Communication we extend previous studies to include both wave and
finite deformation radius effects, so that the model becomes more relevant to
geophysics and plasma physics applications and reveals new physical effects.
It is known from observations that, in planetary atmospheres, there exist
global-scale circulations in the form of longitudinal zonal flows with
embedded long-lived vortices. Indeed, this effect is significant in the
atmospheres of giant planets and in the Earth's oceans. The mechanism of the
formation of zonal flows on the surface of giant planets is a separate
question which is not well understood. Two distinct approaches to this
problem are based respectively on 3D thermal convection \cite{busse} and on
the inverse energy cascade in 2D turbulence. Very recent data obtained by the
Galileo probe during the unique first-ever {\em in situ} measurements in the
atmosphere of Jupiter \cite{sci} seem to be more in favor of the 3D
convection concept: the observations showed stronger than previously assumed
winds (up to 200 m/s) and turbulence in the upper layer of the Jovian
atmosphere. This indicates that the origin of Jupiter's winds and circulation
patterns is probably heat escaping from Jupiter's deep interior. However, the
eddies we are interested in here are believed to be quasi-2D structures
originated from the instability of zonal flows and confined to a shallow
atmospheric layer. 

There have been several studies devoted to the combined impact of finite
deformation radius and the $\beta$-effect \cite{cush,pol94}. However, it is
difficult to detect any particular new physical effects, since the scales
overlap thus making the whole picture vague. If the rotation is strong
enough, zonal flows destroy coherent vortices. In the opposite case, when the
deformation radius is smaller than the Rhines length, the formation of
``shielded'' weakly interacting vortices significantly reduces the inverse
cascade, so that zonal flows do not form. We will show how this difficulty
can be partly overcome if the simplistic model is modified to take into
account additional effects that take place when scales of the order or larger
than the deformation radius are considered. 

We consider the generalized geostrophic (or Charney-Hasegawa-Mima) equation
for Rossby-wave turbulence in the $\beta$-plane approximation: 

\begin{equation}
\partial_t {(\nabla^2 h-\lambda^2 h)} + [h,\nabla^2 h]
+\beta \partial_x h (1+h)= D+F,
\end{equation}

\vspace*{0.2mm}

\noindent
where $\lambda$ is the ratio of a spatial scale used for normalization to the
Rossby radius $L_R={(gH_0)}^{1/2}/f_0$, $g$ is the gravitational
acceleration, $h=(H-H_0)/H_0$ is the perturbation of the atmosphere of
average depth $H_0$, $f=f_0+\beta y$ is the Coriolis parameter,
$[a,b]=a_xb_y-a_yb_x$, $D$ and $F$ is the dissipation and forcing
respectively, and the following dimensionless variables are used:  $x,y
\rightarrow \lambda L_R x,~\lambda L_R y, ~~ t \rightarrow \lambda^2 t/f_0,
~~ \beta \rightarrow f_0\beta/\lambda^3 L_R$. Eq.(1) is isomorphic to the
equation for drift waves in a magnetized plasma where the inhomogeneity of
the Coriolis parameter is replaced by the gradient of the electron density
(see, e.g., \cite{horton} for details). Consequently all results can be
transferred to the plasma case. The equation containing the quadratic
$\beta$-term was derived for the first time in \cite{petv80} and has been
studied since then for various applications \cite{nezlin}. The presence of
the scalar nonlinearity $\beta h \partial_x h$ is known to introduce
cyclone/anticyclone asymmetry in Eq.(1) (in a cyclone the direction of
rotation coincides with that of the system). This term appears due to the
perturbation of the fluid depth under the influence of the Rossby wave, so
that the full depth $H=H_0+h$ should be retained rather than the mean depth
$H_0$. 

Studies of shear flows formed due to the strong rotation in $\beta$-plane 2D
models indicate that these flows are stable as described by the Rayleigh-Kuo
instability criterion \cite{vallis,chek}. This result obviously contradicts
the observed coexistence of coherent long-lived vortices and zonal jets in
the Jovian atmosphere. Following \cite{nezlin} we show that this
contradiction can be resolved when the finite deformation radius is taken
into account. The Rayleigh-Kuo criterion for shear flow instability in the
approximation of a constant depth of the atmosphere ($H_0=const$) is given by
${{\partial^2 U} / {\partial y^2}} - \beta_0 = 0$, where $U$ is the average
horizontal flow velocity, $y$ is the meridional coordinate, and
$\beta_0={\partial f / \partial y}$. However, if there is a free surface, the
average depth of the atmosphere $H_0$ can change in the meridional direction: 
${\partial H_0 / \partial y} \neq 0$. This gradient is balanced by the
Coriolis force $g \partial H_0 / \partial y = -fU$, and $\beta$ is given by

\begin{equation}
\beta = - {1 \over {L_R^2}} {\partial \over \partial y} (f L_R^2) =
{\partial f \over \partial y} - {f \over {H_0}} {\partial H_0 \over 
\partial y} =\beta_0 + {U \over {L_R^2}}
\end{equation}

\noindent
Consequently, the modified instability criterion is \cite{nezlin}

\begin{equation}
{{\partial^2 U} \over {\partial y^2}} - \beta_0 - {U \over {L_R^2}}= 0
\end{equation}

\noindent
or in our dimensionless units

\begin{equation}
{{\partial^2 U} \over {\partial y^2}} - \beta_0 - \lambda^2 U = 0
\end{equation}

\noindent
It is now clear that zonal flows which otherwise would be stable can become
unstable when the finite deformation radius is taken into account. At the
same time, the scalar nonlinearity in Eq.(1), as in the case of the
Korteweg-de Vries equation, can lead to the gradual steepening of the initial
perturbation in the longitudinal direction which is compensated by negative
dispersion only for anticyclones.  One can thus expect that vortices emerging
due to the instability of zonal flows will be mostly anticyclones. 

We used the following setup for the numerical experiment to check this
prediction. In 2D turbulence it has been convincingly demonstrated that the
case in which long-term dynamics does not depend on initial conditions is
best achieved with small-scale random forcing. In this case zonal flows are
known to be a robust feature of the flow evolution on the $\beta$-plane.  We
solve Eq.(1) numerically using a pseudospectral method in a square domain $2
\pi \times 2 \pi$ with doubly periodic boundary conditions and resolution
$512 \times 512$. To confine the dissipation to the smallest scales we use
hyperviscosity $D=(-1)^{p+1}\nu_p \nabla^{2p}(\nabla^2 h), ~~ p=8$. We start
with zero initial conditions, random forcing at $100<k_f<105$, and
$\beta=100$, $\lambda = 0$. One can expect the formation of zonal flows to
begin when the inverse energy cascade reaches the Rhines scale $L_{\beta}
\simeq k_{\beta}^{-1}$, $k_{\beta} = {(\beta^2/\epsilon)}^{1/5}$, where
$\epsilon$ is the energy injection rate, and in the considered case
$k_{\beta} \simeq 60$. The flow is dominated by well-developed zonal flows by
$t=15\tau_0$, where $\tau_0 \simeq \pi/v_{rms}$ is the large eddy turnover
time. At this moment the force is turned off to let the viscosity smooth out
the irregularities in the smallest scales. At time $t=17\tau_0$ one can
observe quite regular shear flows. The horizontal velocity has a profile
resembling the one on Jupiter with smooth westward and peaked eastward flows
(Fig. 1a). These flows are stable, since $U_{yy} - \beta_0 < 0$ for all $y$
(Fig. 1b). After that we ``turn on'' the deformation radius (or equivalently
remove the ``rigid lid'' and allow the fluid to have a free surface) by
taking $\lambda =10$. The choice of $\lambda$ is made such that the
parameters would be close to the ones for Jupiter's Great Red Spot (JGRS). 
For Jupiter ${R / L_R} \simeq {{7 \cdot 10^4 km} / {6 \cdot 10^3 km}} \sim
10$, $f_{GRS} \simeq 1.4 \cdot 10^{-4} s^{-1}$, and since $\beta \sim f/R$,
this gives $\beta \simeq 100$ in our dimensionless units. The value $U_{yy} -
\beta_0 -\lambda^2U$ now changes sign for some $y$, and the westward
anticyclonic ($U<0$) zonal flow become unstable. This almost immediately
($\Delta t \sim \tau_0$) leads to the formation of vortices. They presumably
appear as cyclone-anticyclone pairs with cyclones decaying faster, so that
only anticyclones survive (Fig.2). To illustrate the asymmetry of the
solution, in Fig.3 we plot the vorticity skewness
$S_{\omega}(t)={\langle{\omega}^3\rangle}$/${~\langle{\omega}^2\rangle}^{3/2}$,
where $\omega=\nabla^2h$, $\langle~~\rangle$ denotes an area average ($\omega
< 0$ corresponding to anticyclones). 

Another important feature to verify is the form of the vortices. As was
discussed in \cite{petv80,flierl} the scalar nonlinearity in Eq.(1) allows
soliton-like solutions. These solutions combine characteristic features of
both waves and real vortices, in the sense that they have trapped fluid, so
they are called ``solitary vortices''. An analytically obtained solution
\cite{petv80} under the assumption $h \ll H_0$ predicts the structure of such
a soliton to be given by

\begin{equation}
h/h_0=cosh^{-3/4}({3r \over 4a}), ~~~~a \simeq L_R h_0^{-1/2}
\end{equation}

\noindent
Consequently, the vorticity profile $\omega=\nabla^2 h={1 \over r}{\partial
\over \partial r}(r{\partial h \over \partial r})$ is smooth, gradually
decreasing (increasing) from the center of a vortex. Observations show,
however, that, for example, the JGRS and intrathermoclinic vortices (lenses)
in the Earth's ocean do not rotate as solid bodies, and their vorticity has a
circumferential profile with a rotating perimeter and relatively quiet core.
Also in experiments in rotating tanks \cite{antipov} with $h \simeq O(1)$,
vortices appear to be more localized than predicted by the theoretical
solution (5). Sutyrin noted \cite{sut} that much better agreement with
experiments can be obtained if one uses the assumption that there is a
uniform profile of the potential vorticity $\xi=\nabla^2 h-\lambda^2h=const$
inside the vortex, although the vorticity profile was never analyzed.
Ironically, Marcus \cite{marcus} used the same assumption to criticize the
soliton-like model, arguing that it is unable to produce ring vortices. 

We show here that the model Eq.(1) does produce ring vortices and we present
the evidence that it is the scalar nonlinearity that plays a decisive role. 
It is important to emphasize that these vortices spontaneously arise from
random initial conditions; they are not in any sense ``trial'' vortices
introduced into the flow. 

In our numerical experiment without forcing we start with the following
initial perturbation

\vspace{-0.2in}
\[ h_{\bf k} = \left\{ \begin{array}{ll}
                              Cexp(i\phi_{\bf k}) & \mbox{if ~$4<k<5$ }\\
                   0 &  \mbox{otherwise,}
                   \end{array} \right. \]

\noindent
with random phases $\phi_{\bf k} \in [0, 2\pi)$. $C$ is chosen such that
total initial energy, $E={1 \over 2}\sum_k (k^2+\lambda^2){\vert h_{\bf
k}\vert}^2$, is equal to 0.5, and $\lambda=L_R^{-1}=10,~~\beta=20$. The
characteristic size of the initial eddies is $L \simeq (6 \div 7)L_R$. During
the evolution of the system for over $50\tau_0$ we observe the formation of
elongated vorticity sheets, which deform into loops. These loops could close
into ring vortices, all of them being anticyclones (Fig.4). If the scalar
nonlinearity is dropped from the equation, closed loops and ring vortices (in
this case they could equally be cyclones and anticylones) do not form
(Fig.5a).  In the case $\beta=0$ the cubic nonlinearity $J(h,h\nabla^2h+{1
\over 2}(\nabla h)^2)$ should be retained in Eq.(1) \cite{cush}.  The
cyclone/anticyclone asymmetry is then preserved, but no signs of ring
vortices are found (Fig.5b). 

We do not have a theoretical explanation of the formation of ring vortices,
but necessary physical conditions can be identified. First of all, because of
the finite deformation radius $L_R$ vortices are shielded, i.e. their
interaction decreases exponentially. This means that a ring vortex should be
of the size of several $L_R$ in order to be stable. Simultaneously, the
Rhines length $L_{\beta}$ and the Rossby radius $L_R$ should be comparable;
in the opposite case either zonal flows or ``frozen'', not mobile, vortices
dominate the flow. Mutual action of negative dispersion and KdV-like scalar
nonlinearity can lead to the formation of loops. Connection of the opposite
sides of a loop leads to the formation of a ring vortex. Thus, both finite
deformation radius and the scalar nonlinearity are crucial for the emergence
and stability of ring vortices. 

In conclusion, we have studied the question determining the combined
influence of the $\beta$-effect (both Rossby waves and KdV nonlinearity) and
deformation radius on coherent structures.  It is shown that for a certain
set of parameters (close to those of giant planets) large-scale zonal flows
created by small-scale forcing tend to preferentially form anticyclonic
vortices. Their emergence can be explained by the modified Rayleigh-Kuo
instability criterion, which takes into account the deformation of free
surface. We have provided numerical evidence that for scales larger than the
Rossby radius, the scalar nonlinearity is responsible for the formation of
ring anticyclonic vortices, a feature known from observations. We have
demonstrated how gradual complication of the simple model allows one to
clearly reveal new physical effects. The complete and accurate study of these
effects will need simulations of more complicated equations on the sphere (in
the spirit of \cite{amar,polv96}). 

This work was partially supported by ARPA/ONR Grant N00014-92-J-1796.

\section*{Figure captions}

\noindent
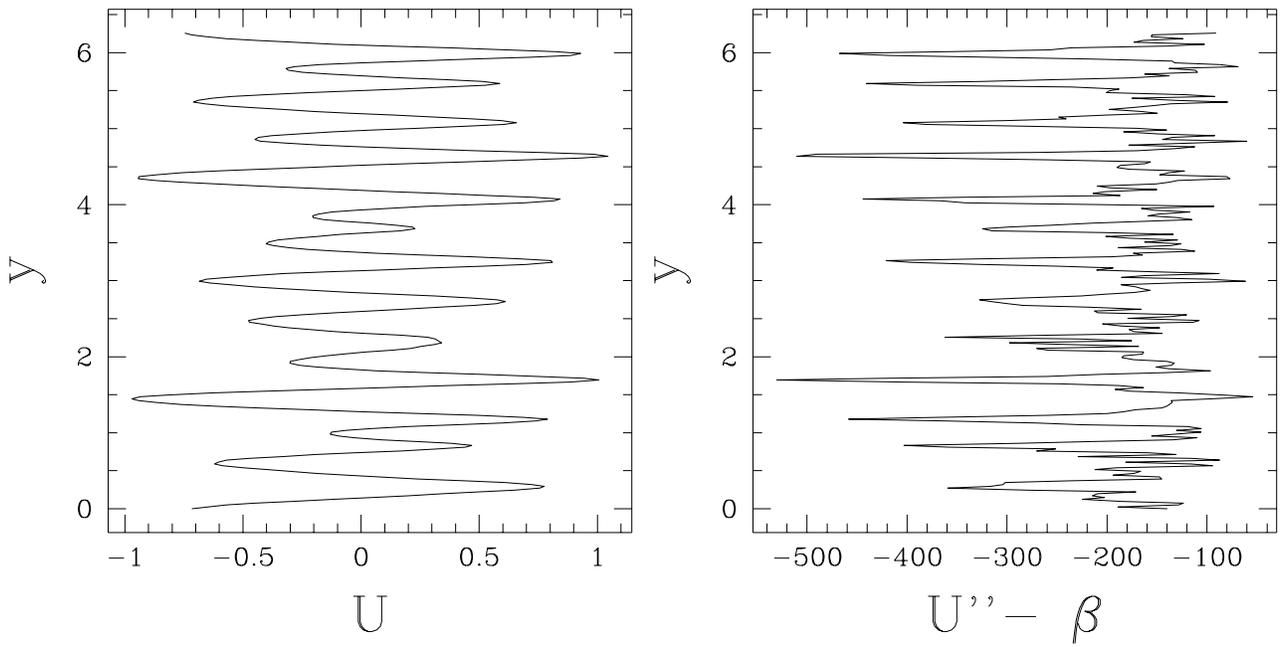
{\em Figure 1.} Averaged zonal velocity $U$ (a) and $U_{yy}-\beta$ (b) as
functions of latitude $y$. 

\vspace* {0.1 in}

\noindent
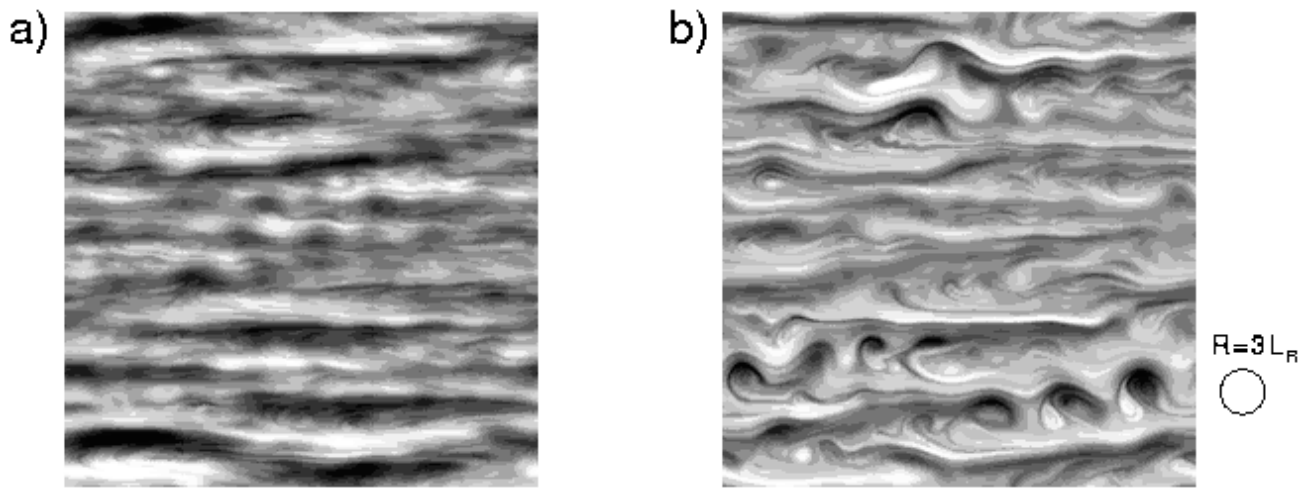
{\em Figure 2.} Finite deformation radius makes zonal flows unstable with
respect to generation of anticyclones. The vorticity field before (a) and
after (b) introducing a finite deformation radius $L_R$ (whose scale is shown
in the right picture). Darker colors correspond to the anticyclons. 

\vspace* {0.1 in}

\noindent
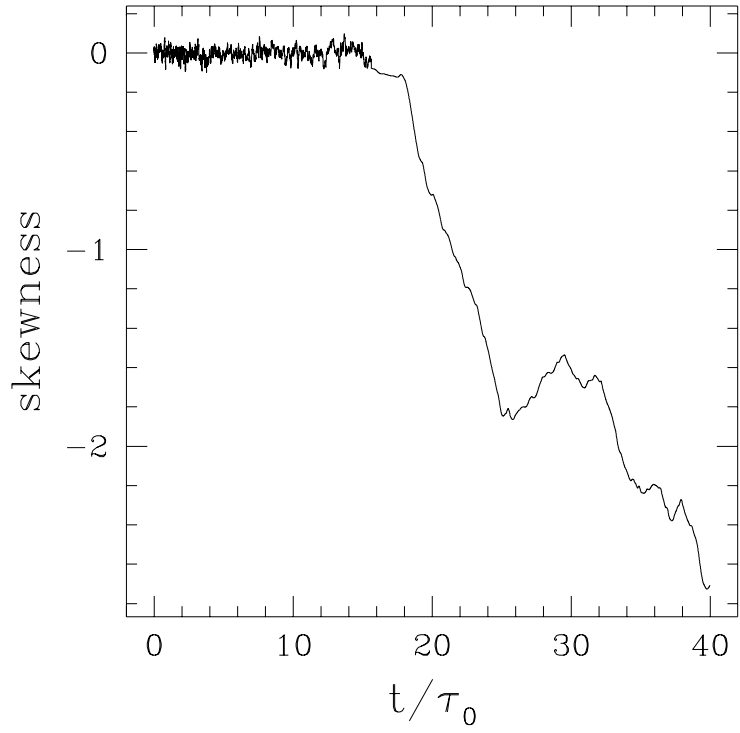
{\em Figure 3.} Vorticity skewness $S_{\omega}(t)$, $\lambda=0$ before
$t=15\tau_0$ and $\lambda=10$ afterwards. 

\vspace* {0.1 in}

\noindent
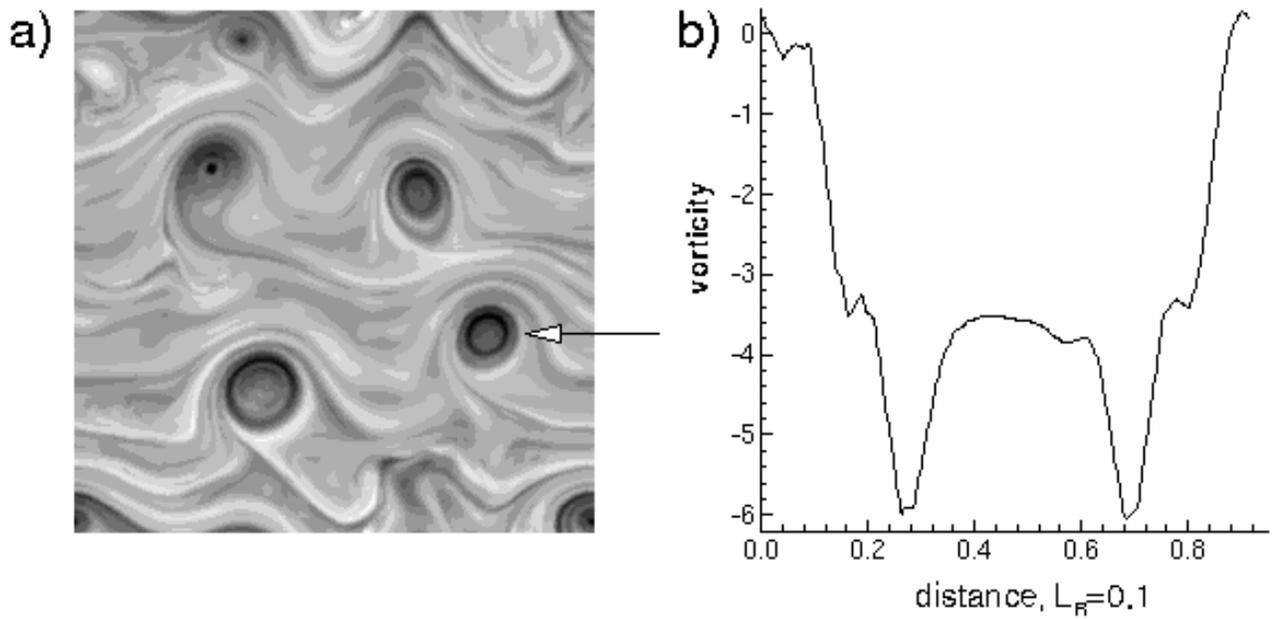
{\em Figure 4.} Instantaneous vorticity field at $t=40\tau_0$ showing the
ring anticyclonic vortices, $S_{\omega}=-2.8$ (a); cross section of a typical
vortex (b). 

\vspace* {0.1 in}

\noindent
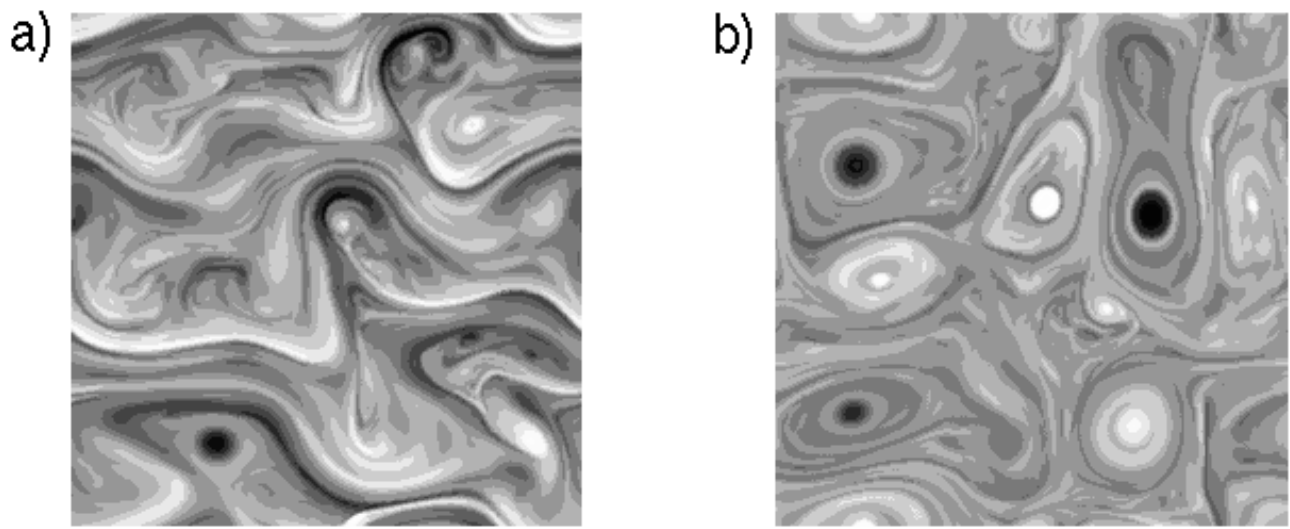
{\em Figure 5.} Vorticity fields at the same time as in Fig.4 when (a) the
scalar nonlinearity is dropped, $S_{\omega}=0$; (b) $\beta=0$, but the term
$J(h,h\nabla^2h+{1 \over 2}(\nabla h)^2)$ is added, $S_{\omega}=-0.3$. 

\end{document}